Quantized representation of some nonlinear integrable evolution equations on the soliton sector

Yair Zarmi
Jacob Blaustein Institutes for Desert Research
Ben-Gurion University of the Negev
Midreshet Ben-Gurion, 84990, Israel

**Abstract**

The Hirota algorithm for solving several integrable nonlinear evolution equations is suggestive of a simple quantized representation of these equations and their soliton solutions over a Fock space of bosons or of fermions. The classical nonlinear wave equation becomes a nonlinear equation for an operator. The solution of this equation is constructed through the operator analog of the Hirota transformation. The classical $N$-solitons solution is the expectation value of the solution operator in an $N$-particle state in the Fock space.

Whereas the structure of the single-soliton solutions of the KdV equation [1],

$$u_t = 6uu_x + u_{xxx} , \tag{1}$$

is simple, the structure of its multiple-solitons solution is rather cumbersome. However, using the Hirota algorithm [2],

$$u(t,x) = 2\partial_x^2 \log[f(t,x)] , \tag{2}$$

the function $f(t,x)$ may be given a simple physical interpretation. For an $N$-solitons solution, with soliton wave numbers $k_i$ $1 \leq i \leq N$, all different from one another, it is given by

$$f(t,x) = 1 + \sum_{i=1}^{N} \varphi(k_i;t,x) + \sum_{n=2}^{N}\left(\sum_{1<i_1<\cdots<i_n}\left\{\prod_{j=1}^{n}\varphi(k_{i_j};t,x)\prod_{i_l<i_m}V(k_{i_l},k_{i_m})\right\}\right)$$
$$\left(\varphi(k;t,x) = e^{2k(x+v(k)t)} , \quad v(k) = 4k^2 , \quad V(k_1,k_2) = \left(\frac{k_1-k_2}{k_1+k_2}\right)^2\right) . \tag{3}$$

As one has $V(k,k') \leq 1$, $f(t,x)$ is bounded by

$$f(t,x) \leq 1 + \sum_{n=1}^{N}\frac{1}{n!}\left(\sum_{i=1}^{N}\varphi(k_i;t,x)\right)^n \leq e^{\left(\sum_{i=1}^{N}\varphi(k_i;t,x)\right)} . \tag{4}$$

Eq. (3) looks like a sum of Feynman diagrams containing single- and multi-particle contributions of all possible subsets of $n \leq N$ of the particles. The functions $\varphi(k_i;t,x)$ may be viewed as real "plane waves", and $V(k,k')$ may be viewed as a "two-particle coupling coefficient".

With this observation in mind, Eq. (3) is suggestive of the following simple quantized representation of the solution of Eq. (1) over a Fock space of bosons or of fermions, with creation and annihilation operators, $a_k^\dagger$ and $a_k$, respectively, and number operators $N_k$, defined by

$$N_k = a_k^\dagger a_k , \quad \begin{pmatrix}[a_k,a_{k'}^\dagger] = \delta(k-k') & (Bosons) \\ \{a_k,a_{k'}^\dagger\} = \delta(k-k') & (Fermions))\end{pmatrix} . \tag{5}$$

Consider the following operator

$$F(t,x) = 1 + \int_0^\infty \varphi(k;t,x) N_k \, dk$$
$$+ \sum_{n=2}^\infty \frac{1}{n!} \int_0^\infty \int_0^\infty \cdots \int_0^\infty \left\{ \left(\prod_{i=1}^n \varphi(k_i;t,x) N_{k_i}\right) \left(\prod_{1 \le l < m \le n} V(k_l, k_m)\right) \right\} dk_1 \, dk_2 \cdots dk_n \quad (6)$$

As $V(k_l, k_m) \le 1$, integration down to $k = 0$, does not pause any problem. To improve the convergence properties of the integrals one may multiply $\varphi(k;t,x)$ by a function of $k$ that falls off sufficiently fast as $k \to \infty$, e.g.,

$$\varphi(k;t,x) \to \varphi(k;t,x) e^{-\alpha k^4} \quad (\alpha > 0) \; . \quad (7)$$

(This amounts to a mere phase shift in the trajectory of a soliton.) For any state with a finite number of particles, the matrix element of the operator $F(t,x)$ is a finite sum of finite terms. Hence, after calculating the matrix element, one may set $\alpha$ to zero.

As $V(k,k') \le 1$, a majorant to the operator $F(t,x)$ is the following operator, the matrix elements of which in states with a finite number of particles are the upper bounds in Eq. (4):

$$M = e^{\int_0^\infty \varphi(k;t,x) N_k \, dk} \; . \quad (8)$$

Denoting a state with $r$ particles with a given wave number, $q$, by $|\{q,r\}\rangle$ (for fermions, obviously, only $r = 1$ is possible), the matrix element of $F(t,x)$ in a single-particle state is:

$$\langle \{q,1\} | F(t,x) | \{q,1\} \rangle = 1 + \varphi(q;t,x) \; . \quad (9)$$

Eq. (9) is identical to the expression for $f(t,x)$ of Eq. (2), when $u(t,x)$ is a single-soliton solution of Eq. (1) [2]. Similarly, the matrix element in a state of two particles with different wave numbers is identical to the expression for $f(t,x)$ when $u(t,x)$ is a two-solitons solution:

$$\langle\{q_1,1\},\{q_2,1\}|F(t,x)|\{q_1,1\},\{q_2,1\}\rangle =$$
$$1+\varphi(q_1;t,x)+\varphi(q_2;t,x)+\varphi(q_1;t,x)\varphi(q_2;t,x)V(q_1,q_2) \quad (10)$$

Extension to $N > 2$ is straightforward: $\langle\{q_1,1\},\cdots,\{q_N,1\}|F(t,x)|\{q_1,1\},\cdots,\{q_N,1\}\rangle$, with $q_i \neq q_j$ ($1 \leq i,j \leq N$, $i \neq j$), is the expression for $f(t,x)$ corresponding to an $N$-soliton solution of Eq. (1).

If the particles are bosons, then a given momentum state may be occupied by more than one particle. A matrix element in a state, in which a given wave number is occupied by several bosons yields a soliton solution with a simple phase shift. For example, the matrix element

$$\langle\{q,n_q\}|F(t,x)|\{q,n_q\}\rangle = 1 + n_q\,\varphi(q;t,x) = 1 + \varphi(q;t,x+\delta) \quad (\delta = \log[n_q]/q) \,, \quad (11)$$

is equal to $f(t,x)$ for a single-soliton solution with a phase shift $\delta$ in the soliton trajectory. The same applies to a state with several distinct wave numbers. For every wave number that is occupied by more than one boson, the corresponding soliton is subjected to a similar phase shift.

The fact that the expression for $f(t,x)$ in the classical case is obtained as the expectation value of a quantum-mechanical operator leads directly to an operator-version of Eq. (1). Consider the operator-analog of Eq. (2):

$$U(t,x) = 2\partial_x\left(F(t,x)_x F(t,x)^{-1}\right) \,. \quad (12)$$

As $F(t,x)$ is a diagonal operator, $U(t,x)$ obeys Eq. (1) on any state with a finite number of particles, and the $N$-soliton-solution of Eq. (1) is equal to the expectation value:

$$u(t,x) = \langle\{q_1,1\},\cdots,\{q_N,1\}|U(t,x)|\{q_1,1\},\cdots,\{q_N,1\}\rangle \,, \quad (13)$$

Using Eq. (12), one can construct operators for the infinite sequence of conserved quantities that characterize the soliton solutions of Eq. (1), as well as for the Hamiltonian, from which Eq. (1) can be derived. For example, the operator corresponding to the first conserved quantity,

$$c_1 = \int_{-\infty}^{+\infty} u(t,x)dx \ , \tag{14}$$

is

$$C_1 = \int_{-\infty}^{+\infty} U(t,x)dx \ . \tag{15}$$

Its action on any state yields

$$C_1 | \{q_1,1\},\cdots,\{q_N,1\}\rangle = c_1(q_1,\cdots,q_N) \ , \tag{16}$$

where $c_1(q_1,\ldots, q_N)$ is the value of $c_1$ for the corresponding $N$-solitons solution.

The same ideas apply to several other integrable equations.

Sawada-Kotera equation [3, 4]

$$u_t = 45 u^2 u_x + 15 u u_{xxx} + 15 u_x u_{xx} + u_{xxxxx} \ . \tag{17}$$

Eq. (17) is integrable [3, 4]. Its soliton solutions are also given by Eqs. (2) and (3), with

$$V(k,k') = \left(\frac{k-k'}{k+k'}\right)^2 \left(\frac{k^2 - kk' + k'^2}{k^2 + kk' + k'^2}\right) \ . \tag{18}$$

mKdV equation [5, 6]

$$u_t = 6 u^2 u_x + u_{xxx} \ . \tag{19}$$

Eq. (19) is integrable [5-7]. Its soliton solutions are given by [7]

$$u(t,x) = 2 \partial_x \tan^{-1}(g(t,x)/f(t,x)) \ . \tag{20}$$

In Eq. (20),

$$g(t,x) = \sum_{i=1}^{N} \varphi(k_i;t,x) + \sum_{\substack{n=3 \\ n \text{ odd}}}^{N} \left( \sum_{1 < i_1 < \cdots < i_n} \left\{ \prod_{j=1}^{n} \varphi(k_{i_j};t,x) \prod_{i_l < i_m} V(k_{i_l}, k_{i_m}) \right\} \right) \ , \tag{21}$$

$$f(t,x) = 1 + \sum_{\substack{n=2 \\ n \text{ even}}}^{N} \left( \sum_{1 \leq i_1 < \cdots < i_n} \left\{ \prod_{j=1}^{n} \varphi(k_{i_j}; t, x) \prod_{i_l < i_m} V(k_{i_l}, k_{i_m}) \right\} \right), \quad (22)$$

$$V(k_1, k_2) = -\left( \frac{k_1 - k_2}{k_1 + k_2} \right)^2. \quad (23)$$

In this case, corresponding to the functions $f(t,x)$ and $g(t,x)$, there are two operators, $F(t,x)$ and $G(t,x)$, which contain terms with, respectively, even and odd $n$ in Eq. (6).

Bidirectional KdV equation [8-10]

$$u_{tt} - u_{xx} - \partial_x (6 u u_x + u_{xxx}) = 0. \quad (24)$$

Eq. (24) is integrable [8-10]. Its soliton solutions are given by Eqs. (2) and (3). The solitons may move in either direction along the $x$-axis. Hence, their velocities are given by

$$v(k, \sigma) = \sigma 4 k^2, \quad \sigma = \pm 1. \quad (25)$$

In addition, the "coupling coefficients" $V(k,k')$ are replaced by ones, which, depend on the wave numbers, as well as on the velocities. For the scaling employed in Eq. (24), they are given by:

$$V(k, \sigma, k', \sigma') = \frac{12(k - k')^2 + (v(k,\sigma) - v(k',\sigma'))^2}{12(k + k')^2 + (v(k,\sigma) - v(k',\sigma'))^2}. \quad (26)$$

These coefficients vanish in the single-particle limit ($k' = k$, $\sigma = \sigma'$). Therefore, the quantized representation described above can be constructed, with particle states characterized by two "quantum numbers": $k$ and $\sigma$. The fundamental operators are denoted by, $a_{k,\sigma}^\dagger$, $a_{k,\sigma}$ and $N_{k,\sigma}$, and an $N$-particle state – by $|\{q_1, \sigma_1, 1\}, \cdots, \{q_N, \sigma_N, 1\}\rangle$. The operator in Eq. (6) is replaced by

$$F(t,x) = 1 + \sum_{\sigma = \pm 1} \int_0^\infty \varphi(k, \sigma; t, x) N_{k,\sigma} \, dk$$

$$+ \sum_{n=2}^{\infty} \frac{1}{n!} \sum_{i=1}^{n} \sum_{\sigma_i = \pm 1} \int_0^\infty \int_0^\infty \cdots \int_0^\infty \left\{ \left( \prod_{i=1}^{n} \varphi(k_i, \sigma_i; t, x) N_{k_i, \sigma_i} \right) \left( \prod_{\substack{1 \leq l < m \leq n \\ \sigma_l, \sigma_m = \pm 1}} V(k_l, \sigma_l, k_m, \sigma_m) \right) \right\} dk_1 \, dk_2 \cdots dk_n \quad (27)$$

The fact that $\sigma$ has two values is suggestive of a formulation in terms of spin-1/2 fermions.

In classical soliton dynamics, the single-soliton solution plays a unique role. There is an infinite hierarchy of differential polynomials in $u$, the solution of an evolution equation, which vanish identically when $u$ is a single-soliton solution ("special polynomials" [11, 12]). As an example, consider the case of the KdV equation. The lowest scaling weight, in which special polynomials exist, is 3. There are two special polynomials in this scaling weight, given by:

$$R^{(3,1)}[u] = u_x + q^{(1,1)} u \quad , \quad R^{(3,2)} = \frac{3}{2}\left(\int_{-\infty}^{x} q^{(1,1)} R^{(3,1)}[u]dx - \int_{x}^{\infty} q^{(1,1)} R^{(3,1)}[u]dx\right)$$
$$\left(q^{(1,1)} = \frac{1}{2}\left(\int_{-\infty}^{x} u(t,x)dx - \int_{x}^{\infty} u(t,x)dx\right)\right) \tag{28}$$

(In each superscript, $(W,i)$, $W$ is the scaling weight, and $i$ counts the polynomials with this scaling weight.) Replacing in Eq. (28) the function $u(t,x)$ by the operator $U(t,x)$ of Eq. (12), both special polynomials become operators, which project the full Fock space into its multi-particle subspace.

The polynomials in Eq. (28) are non-local; they contain integrals over $x$. (Yet, they are bounded.) A local special polynomial (containing only powers of $u$ and of its spatial derivatives) first appears at scaling weight 6. It is given by [11, 12]

$$R^{(6,1)}[u] = u^2 \partial_x\left(\frac{R^{(3,1)}}{u}\right) = u^3 + u u_{xx} - (u_x)^2 \quad . \tag{29}$$

(There are other special polynomials in this scaling weight. They are all non-local.) Using Eq. (12), one can construct the corresponding projection operator:

$$R^{(6,1)}[U] = U(t,x)^3 + U(t,x)\partial_x^2 U(t,x) - (\partial_x U(t,x))^2 \quad . \tag{30}$$

Again, the action of this operator on any single-particle state is readily found to vanish:

$$R^{(6,1)}[U]|\{q,1\}\rangle = 0 \ . \tag{31}$$

Thus, the special polynomials correspond to an infinite hierarchy of commuting projection operators.

The quantized representation depends crucially on the fact that the "coupling coefficients" in Eq. (3) vanish in the limit $k_i = k_j$, $i \neq j$. Hence, such a representation is not possible if this requirement is not satisfied. Two examples are given in the following.

Kaup-Kupershmidt equation [13,14]

$$u_t = 180\, u^2 u_x + 30\, u u_{xxx} + 75\, u_x u_{xx} + u_{xxxxx} \ . \tag{32}$$

Eq. (32) is integrable [15-20]. Its multiple-soliton solutions are given by

$$u(t,x) = \frac{1}{2}\partial_x^2 \log[f(t,x)] \ . \tag{30}$$

The "plane waves", $\varphi(k;t,x)$, are defined as in Eq. (3), with soliton velocities given by

$$v(k) = 16\, k^4 \ . \tag{33}$$

However, the structure of $f(t,x)$ does not follow the pattern of Eq. (3). For the single-soliton solution one has

$$f(t,x) = 1 + \varphi(q,t,x) + \frac{1}{16}\varphi(q,t,x)^2 \ . \tag{34}$$

In the two-solitons solution, the expression for $f(t,x)$ is:

$$\begin{aligned}f(t,x) &= 1 + \varphi(q_1;,t,x) + \varphi(q_2;,t,x) \\ &+ \frac{1}{16}\varphi(q_1;,t,x)^2 + \frac{2q_1^4 - q_1^2 q_2^2 + 2q_2^4}{2(q_1+q_2)^2(q_1^2+q_1 q_2+q_2^2)}\varphi(q_1;,t,x)\varphi(q_2;,t,x) + \frac{1}{16}\varphi(q_2;,t,x)^2 \\ &+ V(q_1,q_2)\left(\varphi(q_1;,t,x)^2\, \varphi(q_2;,t,x) + \varphi(q_1;,t,x)\varphi(q_2;,t,x)^2\right) + V(q_1,q_2)^2\, \varphi(q_1;,t,x)^2\, \varphi(q_2;,t,x)^2 \\ &\left(V(q_1,q_2) = \frac{(q_1-q_2)^2(q_1^2 - q_1 q_2 + q_2^2)}{16(q_1+q_2)^2(q_1^2 + q_1 q_2 + q_2^2)}\right)\end{aligned} \tag{36}$$

Obviously, not all two-wave "coupling coefficients" vanish in the limit $q_1 = q_2$.

Caudrey-Dodd-Gibbon equation [4],

$$u_t = 420 u^3 u_x + 210 u^2 u_{xxx} + 420 u u_x u_{xx} + 28 u u_{xxxxx} + 28 u_x u_{xxxx} + 70 u_{xx} u_{xxx} + u_{xxxxxxx} \quad . \quad (37)$$

The integrability of Eq. (37) is still an open question. The single- and two-solitons solutions do follow the Hirota structure of Eqs. (2) and (3). The two-particle "coupling coefficient" is [4]:

$$V(k, k') = \left( \frac{k - k'}{k + k'} \right)^2 \left( \frac{k^2 - k k' + k'^2}{k^2 + k k' + k'^2} \right)^2 \quad . \quad (38)$$

Attempting to construct a three-solitons solution of Eq. (37), one finds that the coefficients of the second-order terms, $g(k_i;t,x) \cdot g(k_j;t,x)$ ($1 \leq i, j \leq 3, i \neq j$), are of the Hirota form with $V(k_i, k_j)$ of Eq. (38). However, although the third-order term, $g(k_1;t,x) \cdot g(k_2;t,x) \cdot g(k_3;t,x)$, does vanish if any of the two wave numbers are equal, it cannot be factorized into a product of the two-particle coefficients $V(k_i, k_j)$. The same applies to the (necessary) fourth-order terms.

A very simple quantized representation exists in the case of the Burgers equation [21],

$$u_t = 2 u u_x + u_{xx} \quad . \quad (39)$$

The shock-front solutions of Eq. (39) are obtained through the Forsyth-Hopf-Cole transformation [22-24]:

$$u(t,x) = \partial_x \log[f(t,x)] \quad . \quad (40)$$

Here, $f(t,x)$ has the following simple structure:

$$f(t,x) = 1 + \sum_{i=1}^{N} \varphi(k_i;t,x) \quad \left( \varphi(k;t,x) = e^{k(x + v(k)t)} \quad , \quad v(k) = k \right) \quad . \quad (41)$$

Consequently, the operator $F(t,x)$ is given by:

$$F(t,x) = 1 + \int_{-\infty}^{\infty} \varphi(k;t,x) N_k \, dk \quad . \tag{42}$$

There are many attempts in the literature at rigorous quantization procedures of nonlinear evolution equations [25-39]. The quantized representation of equations and their solutions discussed here is somewhat different. It is characterized by the fact that the coordinates, $t$ and $x$ are mere parameters.

Finally, the proposed quantum-mechanical representation opens a new vista for adding perturbations to a nonlinear wave equation. In classical systems, the perturbation is a functional of the unknown solution, typically, a differential polynomial in the latter. A common way for analyzing the effect of the perturbation is through a Normal Form expansion [40-45, 11]. In this approach, the zero-order approximation is a single-soliton or a multiple-solitons solution of the Normal Form. To this solution, one may apply the quantization procedure delineated above. However, the classical perturbation, as well as the higher-order corrections to the solution in the Normal Form expansion, are then also diagonal operators, functions of the number operator, $N_k$. In the quantum-mechanical version, one may add non-diagonal perturbations, containing terms that will destroy one soliton, and generate another soliton instead, e.g., $a_{k_2}^\dagger a_{k_1}$, or terms that will change the number of solitons, such as $a_{k_3}^\dagger a_{k_2}^\dagger a_{k_1}$.

Acknowledgments Helpful discussions with G.I. Burde and I. Rubinstein are acknowledged.